\documentclass{IEEEtran4PSCC}

\makeatletter
\let\old@ps@headings\ps@headings
\let\old@ps@IEEEtitlepagestyle\ps@IEEEtitlepagestyle
\def\psccfooter#1{%
    \def\ps@headings{%
        \old@ps@headings%
        \def\@oddfoot{\strut\hfill#1\hfill\strut}%
        \def\@evenfoot{\strut\hfill#1\hfill\strut}%
    }%
    \def\ps@IEEEtitlepagestyle{%
        \old@ps@IEEEtitlepagestyle%
        \def\@oddfoot{\strut\hfill#1\hfill\strut}%
        \def\@evenfoot{\strut\hfill#1\hfill\strut}%
    }%
    \ps@headings%
}
\makeatother

\psccfooter{%
        \parbox{\textwidth}{\hrulefill \\ \small{23rd Power Systems Computation Conference} \hfill \begin{minipage}{0.2\textwidth}\centering \vspace*{4pt} \includegraphics[scale=0.06]{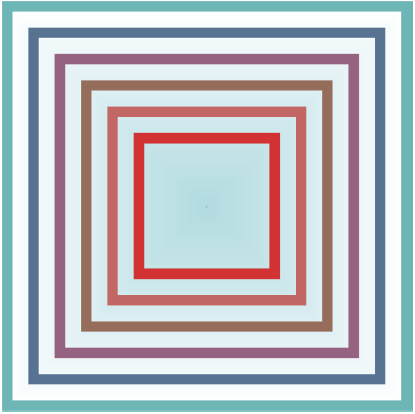}\\\small{PSCC 2024} \end{minipage} \hfill \small{Paris-Saclay, France --- June 4 -- 7, 2024}}%
}

\usepackage{psfrag,color,xcolor,bm,array,cite,bbm}

\usepackage{placeins,float,bbm,xspace,empheq,fancybox,amsmath,amssymb,graphicx,epstopdf,epsfig,subfigure,syntonly,times,amsthm,flushend} 
\usepackage{url,cite,footnote,xspace,syntonly,algorithm,algpseudocode}
\usepackage{verbatim,multirow,slashbox}
\usepackage[T1]{fontenc}
\usepackage{algorithm,cite}
\usepackage{array,booktabs,arydshln}
\usepackage{enumitem,kantlipsum,accents}

\usepackage{tikz}
\usetikzlibrary{positioning}

\newcommand\Algphase[1]{%
\vspace*{-.7\baselineskip}\Statex\hspace*{\dimexpr-\algorithmicindent-2pt\relax}\rule{0.5\textwidth}{0.4pt}%
\Statex\hspace*{-\algorithmicindent}\textbf{#1}%
\vspace*{-.7\baselineskip}\Statex\hspace*{\dimexpr-\algorithmicindent-2pt\relax}\rule{0.5\textwidth}{0.4pt}%
}

\newtheorem{remark}{\bfseries Remark}
\input{mysymbol.sty}

\newcommand{\iu}{\mathrm{i}\mkern1mu}

\begin{document}
%
\title{Data-driven Forced Oscillation Localization\\ using Inferred Impulse Responses}

\author{
\IEEEauthorblockN{Shaohui Liu and Hao Zhu}
\IEEEauthorblockA{Chandra Family Department of ECE  \\
The University of Texas at Austin\\
\{shaohui.liu, haozhu\}@utexas.edu}
\and
\IEEEauthorblockN{Vassilis Kekatos}
\IEEEauthorblockA{Elmore Family School of ECE  \\
Purdue University\\
kekatos@purdue.edu}
}

\maketitle

\begin{abstract}
Poorly damped oscillations pose threats to the stability and reliability of interconnected power systems. In this work, we propose a comprehensive data-driven framework for inferring the sources of forced oscillation (FO) using solely synchrophasor measurements. During normal grid operations, fast-rate ambient data are collected to recover the impulse responses in the small-signal regime, without requiring the system model. When FO events occur, the source is estimated based on the frequency domain analysis by fitting the least-squares (LS) error for the FO data using the impulse responses recovered previously. Although the proposed framework is purely data-driven, the result has been established theoretically via model-based analysis of linearized dynamics under a few realistic assumptions. Numerical validations demonstrate its applicability to realistic power systems including nonlinear, higher-order dynamics with control effects using the IEEE 68-bus system, and the 240-bus system from the IEEE-NASPI FO source location contest. The generalizability of the proposed methodology has been validated using different types of measurements and partial sensor coverage conditions. 
\end{abstract}


\thanksto{\noindent {  This work has been supported by NSF, United States of America, under Grants 1751085, 1923221, 2130706, 2150571, and 2150596.}}


\section{Introduction}
Forced oscillations (FOs) such as the 2019 Florida event ~\cite{nerc2021oscillationref} pose threats to the security and reliability of interconnected power systems, as the potential resonance with a natural system mode could induce oscillations within the interconnection. Among the critical challenges in mitigating FOs, determining the source location using Phasor Measurement Unit (PMU) measurements in real-time has attracted significant attention in recent years~\cite{fo_report2023}.

Thanks to widespread installations of PMUs, synchronized measurements with high sampling rates make it possible to develop data-driven FO localization methods. By calculating the energy function, \textit{dissipating energy flow} (DEF)-based methods have been proposed in~\cite{MASLENNIKOV201755,Osipov2022cpsd,chen2013def}, and adopted by the ISO-NE~\cite{maslennikov2021}. However, the success of DEF-based methods relies upon measurements at the FO source generator and presumes accurate system topology information. Recently, a model-enhanced method has been proposed to address the lack of measurements at the FO sources by utilizing a model of linearized grid dynamics~\cite{lesieutre2022model}. Nevertheless, acquiring accurate system parameters remains a challenge for large interconnections. Other data-driven FO localization methods include a Robust Principal Component Analysis (RPCA)-based method, which extracts the low-rank FO components from PMU measurements~\cite{huang2020forced}, and two recent works suggesting system identification approaches~\cite{delabays2022locating,cai2022online}. However, the RPCA method lacks theoretical connections between the power system model and low-rank FO components, as pointed out by~\cite{delabays2022locating}. While the system identification-based approaches are explainable from a theoretical perspective, the requirement of collecting measurements at most buses may be impractical and incur high computational complexity.

To address the aforementioned issues in the state-of-the-art, this work puts forth a purely data-driven FO localization framework, solidly founded upon linear system analysis of the power system in the small-signal regime. This analysis suggests that the FO input can be fitted in the frequency (Fourier) domain with the FO measurements if the system impulse responses between candidate input-output buses are available. To this end, we genuinely consider using the impulse responses recovered from ambient measurements collected during normal operation of the power system. Our earlier works assert that the response recovery is analytically guaranteed for both internal system states at generators, as well as output variables at buses and transmission lines~\cite{huynh2018data,liu2023dynamic}. By \emph{output variables}, we refer to the algebraic variables of the differential-algebraic equations governing grid dynamics, such as terminal voltage angles/frequencies and active line flows.
The analysis conducted in this work suggests that when direct measurements of generator states are not available, the \textit{relative phase shifts} between output data could still be accurately estimated and, subsequently, used to identify the FO source. We also show that the proposed framework can be readily generalized to handle FO scenarios with multiple sources and frequencies. 

A two-stage algorithm is accordingly developed to recover the FO sources. Before the FO events, ambient PMU measurements are collected to infer impulse responses. When an FO event is detected, the recovered responses are used to identify FO sources. Heed that our algorithm requires measurements only at a nearby bus for each possible FO source. Moreover, the algorithm can leverage diverse types of measurements (including line flows) at any location in contrast to energy-based methods such as~\cite{Osipov2022cpsd}. Finally, the proposed algorithm is computationally efficient, as most of the computational burden is on the offline recovery of impulse responses.

The main contribution of this work is two-fold. First, we analyze the FO responses for both generator states and output variables, and formulate localization problems for both cases accordingly. The proposed FO localization framework utilizes the impulse responses recovered using ambient measurements, and builds upon linearized grid dynamics under reasonable assumptions for wide-area interconnections. Second, we develop a computationally efficient data-driven FO localization algorithm that utilizes PMU data collected before and during the FO event. The algorithm requires no knowledge of the actual system model or dynamics parameters and is flexible regarding measurement types and locations.

The rest of this paper is organized as follows. Section~\ref{sec:fo_model} introduces the linear system dynamics and formulates the FO localization problem. Section~\ref{sec:dyn_model} deals with source localization first from generator state FO data using impulse responses recovered from ambient data. It then generalizes results to collecting FO data from output variables and to settings of FOs having multiple sources. A two-stage localization algorithm is developed in Section~\ref{sec:algorithm}. {  Numerical tests on the IEEE 68-bus system and the IEEE-NASPI FO contest dataset} in Section~\ref{sec:tests} demonstrate the effectiveness of the novel algorithm under different input and measurement conditions. The paper is concluded in Section~\ref{sec:conclusion}.

\section{Problem Formulation}
\label{sec:fo_model}

In the vicinity of an operating point, grid dynamics can be approximated by a linearized state-space model ~\cite[Ch.~14]{arthur2000power}:
\begin{align}
\Dot{\bbx} (t) = \bbA \bbx (t) + \bbB \bbu (t).
\label{eq:lti}
\end{align}
Here vector $\bbx \in \mathbb{R}^{m}$ collects all system state variables, such as generators' internal states, while $\bbu \in \mathbb{R}^n$ collects system inputs corresponding to disturbances to nodal power balance. Both $\bbx$ and $\bbu$ represent deviations from the operating point. Matrices $\bbA$ and $\bbB$ are the state matrix and input matrix, respectively. The linearized model in \eqref{eq:lti} approximates well the power system dynamics of interest to our work focusing on the small-signal oscillation regime~\cite{cai2022online}\cite[Ch.~14]{arthur2000power}. Note that \eqref{eq:lti} is a multiple-input multiple-output (MIMO), and linear time-invariant (LTI) system. Given the system inputs $\bbu(t)$, the states $\bbx(t)$ can be fully determined using the system's impulse response with the subscripts here indicating the input and output. For a single input $u_{\ell}(t)$ at location $\ell$, the output $\bbx(t)$ can be found as the convolution of $u_{\ell}(t)$ with the $\ell$-th column of the impulse response matrix $T_{\bbu,\bbx}(t)$. Expressed in the Fourier domain per the convolution theorem, we get that
\begin{align}
\hhatbbx (\xi) = \hat{T}_{u_\ell,\bbx} (\xi) \cdot \hhatu_\ell (\xi),~\forall \ell = 1,\ldots, n
\label{eq:conv_thm}
\end{align}
where the notation $\hat{\cdot}(\xi)$ stands for the Fourier-domain representation of a signal evaluated at frequency $\xi \in \mathbb{R}$.


Forced oscillations (FOs) are sustained oscillations caused by periodic inputs to power systems. Possible FO sources include cyclic loads, malfunctioning equipment, and controller failures; see \cite{nerc2021oscillationref} for more discussions. Without loss of generality, this work considers FO inputs due to generator controller failures. Typically, turbine governor control directly affects the mechanical power balance in the form of perturbing input $u_\ell(t)$. Recent work \cite{chevalier2018mitigating} has found that exciter control can also be modeled with a similar effect on $u_\ell(t)$. Thus, we focus on these two types of FO sources due to generator controls.

If $u_\ell(t)$ is a periodic sinusoidal signal with $F$ frequency components, its Fourier-domain representation becomes
\begin{align}
    \hhatu_\ell (\xi) \triangleq \sum_{j=1}^F \frac{A_j}{2} \left[ e^{\iu\phi_j}\Delta(\xi - {\xi_j}) + e^{-\iu\phi_j}\Delta(\xi + {\xi_j}) \right] 
    \label{eq:periodic_input}
\end{align}
where $\Delta(\cdot)$ denotes the Dirac delta function. Per component $j$, symbols $\xi_j$, $A_j$, and $\phi_j$ denote its frequency, amplitude, and phase angle, respectively. The number of frequency components $F$ is known to be small with a typical bound of $F\leq2$~\cite{nerc2021oscillationref}. Although a single-frequency input could induce multiple harmonics of odd orders, most of these harmonics are likely to lie outside the [0.1,0.8]~Hz range of inter-area oscillation modes. We first focus on single-frequency oscillations and handle the case $F=2$ in Section~\ref{sec:general}. 

For $F=1$, the single frequency $\xi_1$ in \eqref{eq:periodic_input} will be simply denoted by $\xi^*$. If $\xi^*$ is close to the frequency of a system mode, the \textit{resonance condition} could lead to large-amplitude oscillations observed at locations away from the source, which greatly complicates FO localization~\cite{venkatasubramanian16resonance}.
Under resonance, the frequency $\xi^*$ can be easily estimated by selecting the largest frequency component in the recorded data, that is $\xi^* := \arg\max_{\xi} \| \hhatbbx (\xi) \|_2$.

Knowing $\xi^*$ and the system's impulse response, we can now estimate the input mode $\hat{u}_\ell$  for each possible source location $\ell$ using the \textit{least-squares (LS)} error criterion applied to~\eqref{eq:conv_thm}. This way, the task of FO source localization amounts to selecting the index $\ell$ minimizing 
\begin{align}
\min_{1\leq \ell\leq n} ~ \min_{\hat{u}_\ell
\in \mathbb{C}} &\left\|\hat{T}_{u_\ell,\bbx} (\xi^*) \cdot \hat{u}_\ell-\hhatbbx (\xi^*)\right \|_2^2 \;.
\label{eq:fo_localization0}
\end{align} 
For each $\ell$, the inner LS problem enjoys a closed-form solution. This makes it very fast to search for the location with the smallest LS fitting error, as detailed later.

\begin{remark}{\emph{(Generalizability)}}\label{rmk:model_general} 
The linearized system model in \eqref{eq:lti} has been introduced to simplify the analysis; the tests of Section~\ref{sec:tests} are conducted on realistic settings with nonlinear dynamics and controllers in place. In addition, while only generator controllers have been considered as the FO sources, our approach could be extended to other FO types. As pointed out in~\cite{nerc2021oscillationref}, cyclic outputs from loads and batteries can arise from controller issues in their power-electronic interfaces. However, these controllers tend to have much faster timescales, and thus are not likely to resonate with inter-area modes that are of higher interest to the FO problem. Moreover, the assumptions of single-mode and single-source for the FO taken in \eqref{eq:fo_localization0} represent most of FO cases observed in practice. The generalized scenario of FOs with multiple modes and/or sources is discussed in Sec.~\ref{sec:general} and tested later on, as well. 
\end{remark}


\section{FO Localization via Recovered Responses}
\label{sec:dyn_model}
\allowdisplaybreaks

While the system's impulse responses enable an efficient FO localization per~\eqref{eq:fo_localization0}, it can be challenging to obtain them without having an accurate model of the power system and the dynamic parameters of all generators. To tackle this predicament, we leverage an approach to estimate impulse responses using synchrophasor data under ambient conditions alone~\cite{liu2023dynamic,huynh2018data}. The approach recovers impulse responses by cross-correlating ambient PMU data between pairs of system inputs-outputs, and enjoys guaranteed recovery performance under some reasonable assumptions listed below. For the rest of the paper, we use the superscript $\cdot^a$ to denote recorded signals under ambient conditions and the superscript $\cdot^f$ to denote signals recorded under FO conditions.

\begin{assumption}{\emph{(Swing dynamics)}}
Grid dynamics follow the second-order model
$\bbM\ddot{\bbdelta} + \bbD\Dot{\bbdelta} + \bbK\bbdelta = \bbu$, 
with the state including generator rotor angles in $\bbdelta$ and speed (frequency) deviations in $\bbomega=\Dot{\bbdelta}$.
\label{assump0}
\end{assumption}

\begin{assumption}{\emph{(Homogeneous damping)}}
The ratio between inertia and damping constants is homogeneous across generators, namely $\bbD = \gamma \bbM$ with a constant $\gamma>0$. 
\label{assump1}
\end{assumption}

\begin{assumption}{\emph{(Lossless system)}}
Under lossless lines, the power flow Jacobian $\bbK$ is a symmetric Laplacian matrix.
\label{assump2}
\end{assumption}

\begin{assumption}{\emph{(Ambient condition)}}
The ambient system input is a zero-mean white-noise random process with locational variance proportional to the inertia constant, i.e., having covariance matrix $\bbSigma = \alpha \bbM$ with $\alpha>0$.
\label{assump3}
\end{assumption}

While these assumptions are postulated to simplify the analysis, they tend to hold pretty well in large interconnections. Swing dynamics in (AS\ref{assump0}) has been popularly used for small-signal analysis to study key oscillation modes~\cite[Ch.~14]{arthur2000power}. The proportional damping in (AS\ref{assump1}) for approximating the swing dynamics has been considered in~\cite{huynh2018data,jalali2021inferring,Paganini19,liu2023dynamic}, as a simplification of practical damping settings. For high-voltage transmission lines, power losses are negligible leading to (AS\ref{assump2}) even with most common load models like constant-current ones~\cite{osti_1004165}.
Finally, the ambient condition in (AS\ref{assump3}) implies that areas of larger perturbations have more inertia deployed. While this assumption may not be very realistic, our analysis in \cite{huynh2018data,liu2023dynamic} essentially used it to show that all ambient modes of interest are equally excited.  

Under these assumptions, the pair-wise impulse responses can be recovered by cross-correlating the corresponding ambient PMU measurements as formally stated next~\cite{huynh2018data,liu2023dynamic}. 

\begin{lemma}{\emph{(State's Impulse Responses})}\label{lemma:ambient}
Under (AS\ref{assump0})--(AS\ref{assump3}), the impulse responses of rotor angle or frequency are equivalent to the cross-correlation of the related ambient signals: 
\begin{align}
    T_{u_\ell,\omega_k}(t) &= \frac{2\gamma}{\alpha} C_{\omega_\ell^a,\omega_k^a}(t) = - \frac{2\gamma}{\alpha} \frac{d}{dt}  C_{\omega_\ell^a,\delta_k^a}(t)  \label{eq:freq} \\
    T_{u_\ell,\delta_k}(t) &= -\frac{2\gamma}{\alpha} C_{\omega_\ell^a,\delta_k^a}(t) = -\frac{2\gamma}{\alpha} \frac{d}{dt}C_{\delta_\ell^a,\delta_k^a}(t)
    \label{eq:ang_resp} 
\end{align}
for $t\geq 0$. Here $C_{\cdot,\cdot}$ denotes the cross-correlation operator and the subscripts indicate its two arguments. Because $T_{u_\ell,\omega_k}(t)$ is the derivative of $T_{u_\ell,\delta_k}(t)$, each one of them can be recovered from the other using time differentiation or integration.
\end{lemma}

Lemma~\ref{lemma:ambient} asserts that the cross-correlation of ambient data can recover the impulse responses. For simplicity, we will use $C_{\omega^a_\ell,\bbx^a}(t)$  to represent the recovered  $T_{u_\ell,\bbx}(t)$, assuming suitable operations have been taken per Lemma~\ref{lemma:ambient}. The former enables us to establish a data-driven FO localization approach in the following.

\begin{proposition}{\emph{(Data-driven FO Localization)}} \label{prop:locate} 
Using the recovered impulse responses, the FO source can be identified based on the estimated FO frequency $\xi^*$ by solving: 
\begin{align}
\min_{1\leq \ell\leq n} ~~ \min_{\hat{u}_\ell  \in \mathbb{C}} \left\|\hat{C}_{\omega^a_\ell,\bbx^a} (\xi^*) \cdot \hat{u}_\ell -\hat{\bbx}^f (\xi^*)\right\|_2^2.
\label{eq:fo_localization}
\end{align}
\end{proposition}

Proposition~\ref{prop:locate} builds upon the recovered responses as the feature vectors for each possible source. Hence, any recovery accuracy issue faced by the data-driven approach in Lemma~\ref{lemma:ambient} could lead to errors in FO localization, as shown by our numerical tests. Heed that there is an unknown scaling factor of $(2\gamma/\alpha)$ in Lemma~\ref{lemma:ambient}. Nonetheless, this factor is the same amongst all locations under (AS\ref{assump1}) and (AS\ref{assump3}), and can thus, be omitted upon solving~\eqref{eq:fo_localization}. The proposed data-driven localization approach can benefit from the fact that each possible source location would lead to very different system-wide responses. It is also known that the participation factors of an oscillation mode at individual generators can be very different, especially from one area to another~\cite[Sec.~8.3]{sauer2017power}. This renders the proposed FO approach more robust against possible recovery errors as demonstrated in Section~\ref{sec:tests}. In general, the more state variables are observed, the better the localization performance. 

Back to \eqref{eq:fo_localization}, the solution of the inner LS problem is simply:
\begin{align}
\hat{u}_\ell \leftarrow \left[\hat{C}_{\omega^a_\ell,\bbx^a} (\xi^*)\right]^+ \hat{\bbx}^f(\xi^*) \;
\label{eq:ls_local_sol}
\end{align}
where superscript $^+$ denotes the matrix's pseudo-inverse. Accordingly, the LS fitting error for location $\ell$ becomes 
\begin{align*}
\left\| \hat{\bbx}^f (\xi^*) -
\hat{C}_{\omega^a_\ell,\bbx^a} (\xi^*)\left[\hat{C}_{\omega^a_\ell,\bbx^a} (\xi^*)\right]^+ \hat{\bbx}^f(\xi^*)    
\right\|_2^2 \;.
\end{align*}
This closed-form solution of LS makes it very convenient to evaluate the fitting error for each possible input location and thus to search for the best one. 


\subsection{Localization based on Output Observations}

Proposition~\ref{prop:locate} assumes synchrophasor readings of generators' internal states, such as $\bbdelta$ and $ \bbomega$, under both ambient and oscillatory conditions. However, PMUs can only measure algebraic variable outputs at the power system level, at buses and lines. We next extend the proposed approach to consider measurements of system outputs. Specifically, we consider measurements of bus voltage phase angle $\theta_i$ and frequency $f_i$ per bus $i$, as well as the active line flow $p_{ij}$ for the line $(i,j)$ connecting buses $i$ and $j$. Other measurement types such as voltage magnitudes and reactive power could be included.  

By linearizing the power flow at a given operating point, system outputs $\bby$ can be linearly related to system states through an observation matrix $\bbC$ as
\begin{align}
\bby \simeq \bbC \bbx \;
\label{eq:linear_pf}
\end{align} 
where $\bby \in \mathbb{R}^{m'}$ collects all the aforementioned bus and line outputs; recall that the dynamic model in \eqref{eq:lti} was also derived upon linearization. Our earlier work~\cite{liu2023dynamic} has utilized the linearized model \eqref{eq:linear_pf} to extend Lemma~\ref{lemma:ambient} to consider output responses for bus and line variables, as stated next.

\begin{lemma}{\emph{(Output's Impulse Responses)}}\label{lemma:ambient2}
Under (AS\ref{assump0})--(AS\ref{assump3}) and \eqref{eq:linear_pf}, the  impulse response of bus angle $\theta_i$ and frequency $f_i$ have the following equivalence with $t\geq0$:
\begin{align}
T_{u_\ell,f_{i}}(t) &= \frac{2\gamma}{\alpha}{C}_{\omega_\ell^a,f_i^a}(t) \label{eq:bus_freq} \\
T_{u_\ell,\theta_{i}}(t) &= -\frac{2\gamma}{\alpha}{C}_{\omega_\ell^a,\theta_i^a}(t). \label{eq:bus_angle} 
\end{align}
For line flow $p_{ij}$,  it similarly holds that:
\begin{align}
T_{u_\ell,p_{ij}}(t) = -\frac{2\gamma}{\alpha}{C}_{\omega_\ell^a,p_{ij}^a}(t). \label{eq:freq_flow}
\end{align}
\end{lemma}

Similar to Proposition~\ref{prop:locate}, Lemma~\ref{lemma:ambient2} enables us to use the recovered responses for bus angles/frequencies and line flows to localize FOs. Nonetheless, recovering these impulse responses still relies on measurements of state variables, i.e., the input generator speed ${\omega}_\ell^a(t)$ in \eqref{eq:bus_freq}--\eqref{eq:freq_flow}. Therefore, Lemma~\ref{lemma:ambient2} is not practically useful. To tackle this issue, we could use the frequency at the nearest bus for each candidate FO source as a surrogate, i.e., using $f_k^a(t)$ at the bus $k$ that is closest to input $\ell$ to replace $\omega^a_\ell(t)$. This naive approach could expand Proposition~\ref{prop:locate} to using all system output measurements, but may introduce inaccuracies to FO localization in~\eqref{eq:fo_localization}. The nearest-bus frequency $f_k^a(t)$ does not exactly match the generator speed $\omega_{\ell}^a(t)$, because per \eqref{eq:linear_pf} the former is in fact a linear combination of all generator states, not just the closest one. This mismatch has also been confirmed numerically, leading to high FO localization error for the naive approach.

To address this observability predicament, we develop a formal approach to account for the inaccuracy in the recovered responses. Consider the Fourier transform of the cross-correlation $C_{y_i^a, y_j^a}(t)$ between two outputs at buses $i$ and $j$. 
Its Fourier-domain representation is called the \textit{cross power spectral density (CPSD)} and will be denoted by $\hat{C}_{y_i^a, y_j^a}(\xi)$. 
Interestingly, the phase angle of CPSD can be shown to reflect the relative phase shift from $\hat{T}_{u_\ell,y_i}$ to $\hat{T}_{u_\ell,y_j}$ as follows~\cite{Trudnowski2008mode}:
\begin{align}
\angle \hat{C}_{y_i^a, y_j^a} (\xi) 
=  \angle\hat{T}_{u_\ell,y_j}(\xi)-\angle \hat{T}_{u_\ell,y_i}(\xi) \; , \forall \xi .
\label{eq:fo_resp_ang}
\end{align}
Hence, the phase angles of the terms in \eqref{eq:conv_thm} relate as: 
\begin{align}
\angle \hat{y}_j^f(\xi) &= 
\angle \hat{u}_\ell(\xi)  + \angle \hat{T}_{u_\ell, y_j} (\xi) \nonumber\\
 &= 
\angle \hat{u}_\ell(\xi)  + \angle \hat{T}_{u_\ell, y_i} (\xi) + \angle \hat{C}_{y_i^a, y_j^a} (\xi) 
 \;  , \forall \xi .
\label{eq:fo_shift}
\end{align}
Equation~\eqref{eq:fo_shift} asserts that we can use the measured output signal $y_i^a(t)$ as a surrogate for the non-measured candidate input signal $\omega_\ell^a(t)$. Nonetheless, by doing so, we introduce an unknown angle $\angle \hat{T}_{u_\ell, y_i} (\xi)$ for all output FO measurements in $\hat{\bby}$. Because this angle ambiguity is the same for all output locations, it can be eliminated by fitting the angle relations in \eqref{eq:fo_shift} with an unknown input angle, instead of the input frequency component in Proposition~\ref{prop:locate}. In other words, for each candidate location $\ell$ and surrogate location $i$, we can alternatively estimate $\angle \hat{u}_i(\xi)=\angle \hat{u}_\ell(\xi)  + \angle \hat{T}_{u_\ell, y_i} (\xi)$ instead of $\angle \hat{u}_\ell(\xi)$ alone; recall each candidate location $\ell$ is paired up with a single surrogate location $i$. This allows us to establish a data-driven FO localization approach  by altering the LS error fitting procedure as shown below.



\begin{proposition}{\emph{(FO Localization using Output Data)}}
\label{prop:fo_local2}
By using the ambient signal $y^a_i$ at the closest bus $i$ for any input location, the FO source can be identified using the phases of the CPSD terms at the estimated FO frequency $\xi^*$ by solving:
\begin{align}
\min_{1\leq i\leq n} \min_{\angle\hat{u}_i} &\| \angle\hat{u}_i+ \angle\hat{C}_{y_i^a,\bby^a} (\xi^*) - \angle\hat{\bby}^f (\xi^*) \|_2^2 \;.
\label{eq:fo_local_phase}
\end{align}
\end{proposition}
Similar to \eqref{eq:fo_localization}, the inner LS problem in~\eqref{eq:fo_local_phase} features a simple closed-form solution given by 
\begin{align}
    \angle\hat{u}_i \leftarrow \frac{1}{m'}\sum_{j=1}^{m'} \left( \angle\hat{y}_j^f (\xi^*)-\angle\hat{C}_{y_i^a,y_j^a} (\xi^*) \right) \;.
\label{eq:fo_shift_sol}
\end{align}
This easy solution makes it very efficient to search for the best input location. Note that \eqref{eq:fo_local_phase} is based solely on output measurements, recorded under ambient conditions and FO conditions. Moreover, it only requires to identify the surrogate output $y_i$ at bus $i$ for each possible input location. Assuming a reasonable level of PMU deployment, the size of output vector $\bby$ could be sufficiently large to allow for distinguishing each input location. Additionally, it can incorporate other types of outputs at any location, thanks to the flexibility of Lemma~\ref{lemma:ambient2}. Hence, the proposed FO localization method can readily incorporate output measurements. In the numerical tests, we will further demonstrate the flexibility of our approach when output measurements closest to the FO source are unavailable.



\subsection{Multiple FO Sources and Modes}
\label{sec:general}

Although most of the analysis thus far focuses on a single FO source with one oscillation frequency, 
it is possible to extend our results to multiple sources/modes based on the  data-driven framework for recovering output responses. Note that the case of single source with multiple frequencies is straightforward to handle. This is because a frequency-domain analysis of the oscillation output $\hat{\bby}^f$ can estimate the number and accordingly the frequencies of underlying oscillation modes by using a prescribed threshold. However, the case of multiple, unknown number of, source locations could greatly complicate the FO localization, as it would become a combinatorial search over all the possible scenarios. Fortunately, it is practically reasonable to assume at most two input source locations for the FO problem, as the chance of having more than two coincidental sources would diminish to almost zero in practice~\cite{nerc2021oscillationref}. Hence, we consider the extension to at most two FO sources with a total of $F$ estimated oscillation frequencies. Accordingly, each frequency $\xi_j^*$ has been obtained by comparing its component with a prescribed threshold $\mathcal{T}$, such that $\| \hhatbby^f (\xi_j^*) \|_2 > \mathcal{T}$. 

Based on the identified frequencies, we extend \eqref{eq:fo_local_phase} to search over any possible pair of input locations $\{i_1,i_2\}$, by solving
%
\begin{align}
\min_{1\leq i_1,i_2\leq n} \sum_{j=1}^F \min_{\angle \hat{u}_{i_1,i_2}} 
\| \angle \hat{u}_{i_1,i_2}  - \angle \hat{\bby}^f (\xi^*_j&) \notag\\
 + \angle \hat{C}_{y^a_{i_1},\bby^a}(\xi^*_j) 
  + \angle \hat{C}_{y_{i_2}^a,\bby^a}(\xi^*_j)
 \|_2^2 &\;.
\label{eq:fo_loc_multi_m_s}
\end{align}
Clearly, each subproblem still follows the same structure as the inner LS minimization in \eqref{eq:fo_local_phase} with the averaging solution. This allows us to quickly compute the minimal fitting error for a candidate pair  $\{i_1,i_2\}$, and then pick the pair with the best fit. 
For the possibility of two input locations, the number of candidate choices becomes ${n(n-1)}/{2}$ as compared to the $n$ choices for the case of a single source. This increased computational complexity could be addressed by using a more efficient search method based on e.g., sparse signal recovery or compressed sensing \cite{hzgg_tps12}. We do not consider the latter here due to the scope limit but will investigate this computation issue with numerical tests later in Section~\ref{sec:algorithm}.  


\section{FO Localization Algorithm}
\label{sec:algorithm}

\begin{figure}[tb!]
\vspace*{-5mm}
\begin{algorithm}[H]
    \caption{FO Localization}
    \begin{algorithmic}[1]
        \State \textbf{Input:} Ambient data $\bbx^a \in \mathbb{R}^{n\times t1}$, oscillation data~$\bbx^f\in \mathbb{R}^{n\times t2}$, number of sources $n$, passband $\left[ f_1,f_2 \right]$.
        \State \textbf{Output:} The candidate oscillation source $\ell^*$ by LS solution, and set $\mathcal{N}_{\ell^*}$ per graph connectivity.
        \State \textbf{Initialize}: Set the iteration index $i,k=1$, inferred impulse response set $\{C_{x_i^a,x_k^a}\}_{i,k=1}^n \in \mathbb{R}^{t1}$, variables in frequency domain: $\{\hat{C}_{x_i^a,x_k^a}(\xi)\}_{i,k=1}^n \in \mathbb{C}^{t1}$, $\{\hat{x}_i^f(\xi)\}_{i=1}^n \in \mathbb{C}^{t2}$, recovered oscillation input $\hhatbbu \in \mathbb{C}^{n}$, oscillation mode $\xi^*$. 
        \Algphase{Phase 1 - Impulse Responses Inference (off-line)}
        \For{$i$ \textbf{in} range($n$) }
            \State Filter and update $\bbx_i^a$: \newline \-\hspace{0.5cm}  $\bbx^a_i \leftarrow bandpass\left(\bbx_i^a - mean(\bbx_i^a) ,\left[ f_1,f_2 \right] \right)$.
        \EndFor
        \For{$i$ \textbf{in} range($n$) }
        \For{$k$ \textbf{in} range($n$) }
            \State Cross-correlate ambient data streams: $C_{x_i^a  ,x_k^a}$.
            \If{Derivative required for $C_{x_i^a  ,x_k^a}$}
            \State Update $C_{x_i^a  ,x_k^a}$: $C_{x_i^a  ,x_k^a}\leftarrow \frac{d}{dt}C_{x_i^a  ,x_k^a}$.
            \EndIf
        \EndFor
        \State Run FFT and update $\hat{C}_{x_i,\bbx}(\xi)$: $\hat{C}_{x_i,\bbx}(\xi) \xleftarrow{\text{fft}} C_{x_i,\bbx}$.
        \EndFor
        \State \textbf{Return:} inferred impulse response set $\{\hat{C}_{x_i,\bbx}(\xi) \}_{i=1}^n$.
        \Algphase{Phase 2 - FO solver (on-line)}
        \State Run FFT and update $\hat{\bbx}^f$: $\hhatbbx^f(\xi)  \xleftarrow{\text{fft}}  \bbx^f$.
        \State Identify oscillation mode and update $\xi^*$:\newline  $\xi^* \leftarrow \arg\max_{\xi} | \hat{\bbx}^f (\xi) |$.
        \For{$i$ \textbf{in} range($n$) }
        \If{$\bbx^a$ taken at generators}
            \State Solve LS problem and update $\hat{u}_i$: $\hat{u}_i \leftarrow \eqref{eq:ls_local_sol}$.
            \Else
            \State Solve LS problem and update $\angle \hat{u}_i$: $\angle \hat{u}_i \leftarrow \eqref{eq:fo_shift_sol}$.
        \EndIf
        \EndFor
        \State Pick $\ell^*$ with the smallest fitting error, and generate its neighboring set $\mathcal{N}_{\ell^*}$. 
        \State \textbf{Return:} $\ell^*$, $\mathcal{N}_{\ell^*}$.
    \end{algorithmic}
    \label{alg:fo_framework}
\end{algorithm}
\vspace*{-10mm}
\end{figure}

The proposed scheme is implemented as a two-phase algorithm tabulated as Algorithm~\ref{alg:fo_framework}. We consider the realistic condition that PMUs are installed at critical substations with large generation and lines with high power flows connecting control areas. We also assume there is a PMU installed at one nearby bus for each possible FO source generator. During \emph{Phase 1}, pair-wise impulse responses are inferred from ambient measurements. First, we apply a bandpass filter with passband [0.1,0.8]~Hz on ambient data $\bbx^a(\bby^a)$ to find the detrended signals~\cite{nerc2021oscillationref}. For simplicity, both state vector $\bbx$ and output vector $\bby$ are denoted by $\bbx$ in this section. Next, the detrended data streams are cross-correlated for all pairs of input/output locations per Lemmas~\ref{lemma:ambient} and~\ref{lemma:ambient2}. Depending on the type of impulse responses of interest and ambient data, numerical differentiation may be taken to recover $T_{u_\ell,\bbx}$ for all possible input locations $\ell$. The Fast Fourier Transform (FFT) will be taken to generate $\{\hat{T}_{u_\ell,\bbx}(\xi) \}_{\ell=1}^n$.

When an FO event is detected, {\emph{Phase 2}} of the proposed algorithm is initiated to localize the FO source. First, the algorithm runs FFT analysis on $\bbx^f$ to generate $\hat{\bbx}^f(\xi)$, and identify the oscillation frequency $\xi^*$. Next, for each possible $\ell$, the LS solution for input $\hat{u}_\ell$ is calculated using the recovered $\hat{T}_{u_\ell,\bbx}(\xi^*)$ by \eqref{eq:fo_localization0} or \eqref{eq:fo_local_phase} depending on the measurement locations. Finally, the algorithm reports the location $\ell^*$ with the smallest fitting error of~\eqref{eq:fo_localization0} or~\eqref{eq:fo_local_phase}. Note that the neighboring set $\mathcal{N}_{\ell^*}$ of location $\ell^*$ per graph connectivity will also be reported, as the estimation error of recovered impulse responses may lead the identified source to a nearby location.

To show the computational efficiency of the algorithm, we consider {\emph{Phase 1}} and {\emph{Phase 2}} separately. We consider a system with $n$ measurement locations (PMUs) for simplicity. In {\emph{Phase 1}}, the computational complexity is dominated by cross-correlation. Assume each PMU takes $\mathcal{M}_1$ ambient samples. The complexity for each pair of cross-correlation is $\ccalO(\mathcal{M}_1^2)$~\cite{hale2006efficient}. Thus, the complexity of {\emph{Phase 1}} is $\ccalO(n^2\mathcal{M}_1^2)$. 
In {\emph{Phase 2}}, assume each PMU takes $\mathcal{M}_2$ FO samples. The complexity of the FFT step is $\ccalO(n \mathcal{M}_2 \text{log} \mathcal{M}_2)$. Additionally, the complexity of calculating LS solutions and report the FO source is $\ccalO(n^2)$. In general, the number of FO samples $\mathcal{M}_2$ is larger than the number of locations $n$. Thus, the computational complexity of {\emph{Phase 2}} is $\ccalO(n \mathcal{M}_2 \text{log} \mathcal{M}_2)$.

To implement Algorithm~\ref{alg:fo_framework}, we recommend using at least 10~min of ambient data to recover impulse responses, and 20~sec of FO data for localization. The length of recommended measurements indicate that $n \ll \mathcal{M}_2 \ll \mathcal{M}_1$. Thus, the computational burden is mostly at the off-line stage. For on-line FO source localization, {\emph{Phase 2}} is very efficient thanks to the closed-form LS solution, and the small number of required FO samples. Note that for the two-source scenario, the complexity of solving \eqref{eq:fo_loc_multi_m_s} becomes $\ccalO(n^3)$, which does not alter the overall complexity as $n \ll \mathcal{M}_2$.


\section{Numerical Results}
\label{sec:tests}

\begin{figure}[t]
     \centering
         \includegraphics[width=1.0\linewidth]{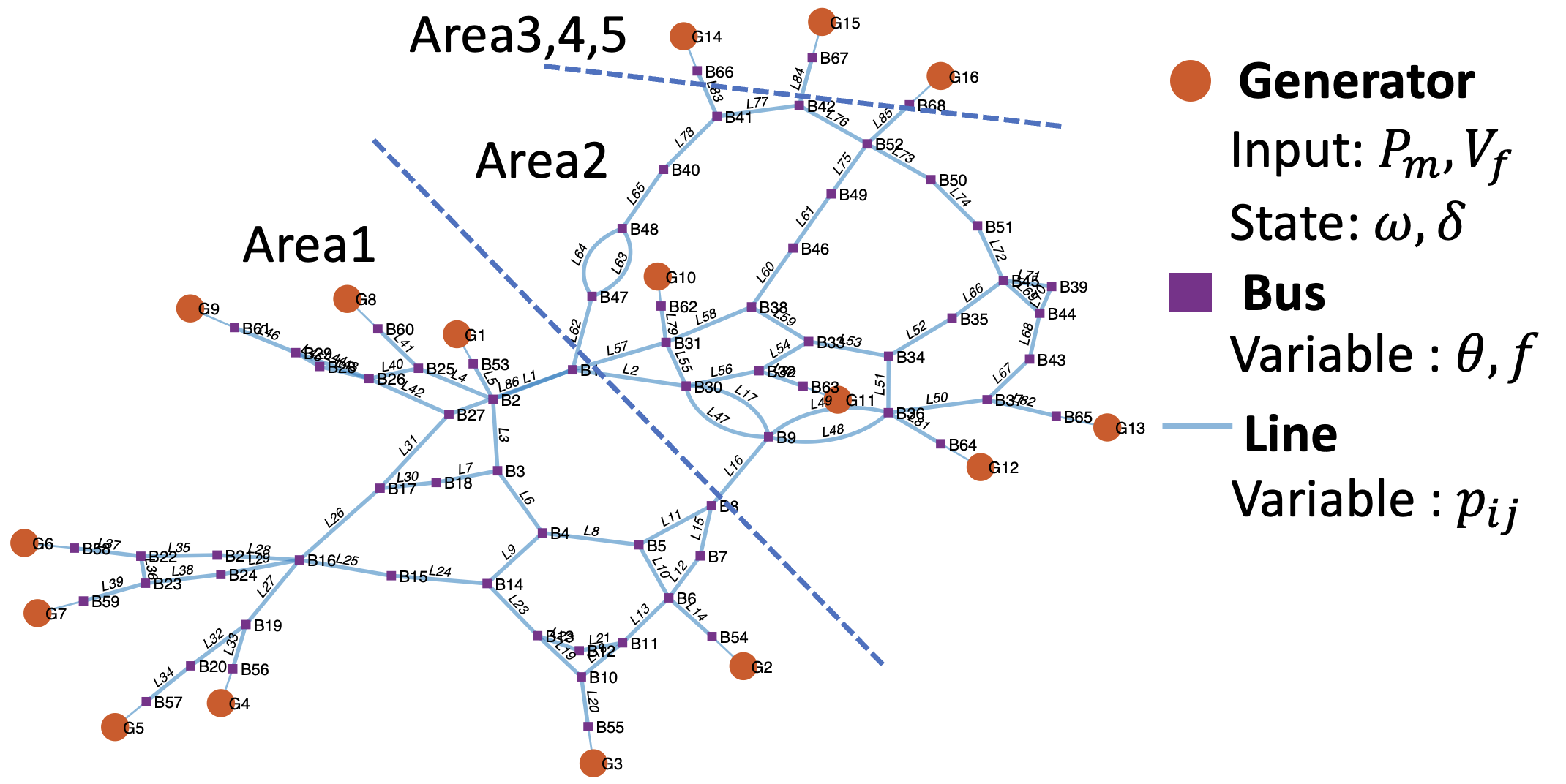}
    \caption{Diagram of the 68-bus system with 16 generators.}
    \label{fig:68bus_diagram}
\end{figure}
{  The proposed method has been numerically evaluated on the IEEE 68-bus system, and the 240-bus system from the IEEE-NASPI oscillation source location
contest~\cite{wang2021ieee}. 
As shown in Fig.~\ref{fig:68bus_diagram}, the 68-bus system has 16 generators, where the 9 generators in Area 1 are equipped with exciters. For this system, impulse responses, ambient data, and forced oscillations were generated in MATLAB using PST~\cite{chow1992pst}, as detailed soon. We considered \emph{all} generators as FO input locations.} We commenced with the single-source, single-FO scenario, with the following cases: \emph{i)} Direct measurements of generator states with full observability; \emph{ii)} Measurements of generators' terminal variables, again with full observability; and \emph{iii)} Measurements only at surrogate buses along with buses and lines as outputs. We subsequently explored the cases of two FO sources and two FO frequencies.\footnote{  Codes and results are available at: \url{https://github.com/ShaohuiLiu/fo_local}.} {  For the 240-bus system, we have used the FO dataset from the IEEE-NASPI contest~\cite{wang2021ieee}, and an ambient dataset generated by the TSAT simulator with random load perturbations for the same system.} 

For the 68-bus system, we simulated nonlinear generator models with control effects and set system inputs at generator mechanical power and exciter voltage references. For impulse responses, we ran time-domain simulations with an impulse-like function (very short rectangular wave?) as input. To generate ambient signals, the mechanical power and exciter references of all generators were perturbed by zero-mean white Gaussian noise. To generate FO signals, the generator's mechanical power or exciter's reference was perturbed by a sinusoidal signal with a frequency close to a system mode. We selected the four modes 0.143Hz, 0.43Hz, 0.57Hz, and 0.714Hz {  (identified as system modes from the simulated impulse responses) as FO input frequencies}. The sampling rate for all tests was set to 200Hz (time-step of $dt=0.005s$). Note that the simulated system is nonlinear; thus, the LTI assumption in~(AS\ref{assump0}) has some mismatch. The line resistance-to-reactance ratio was small, and damping constants $\bbD$ were proportional to inertia constants $\bbM$, making (AS\ref{assump1})--(AS\ref{assump2}) hold pretty well.

\begin{figure}[t]
     \centering
         \includegraphics[width=1.0\linewidth]{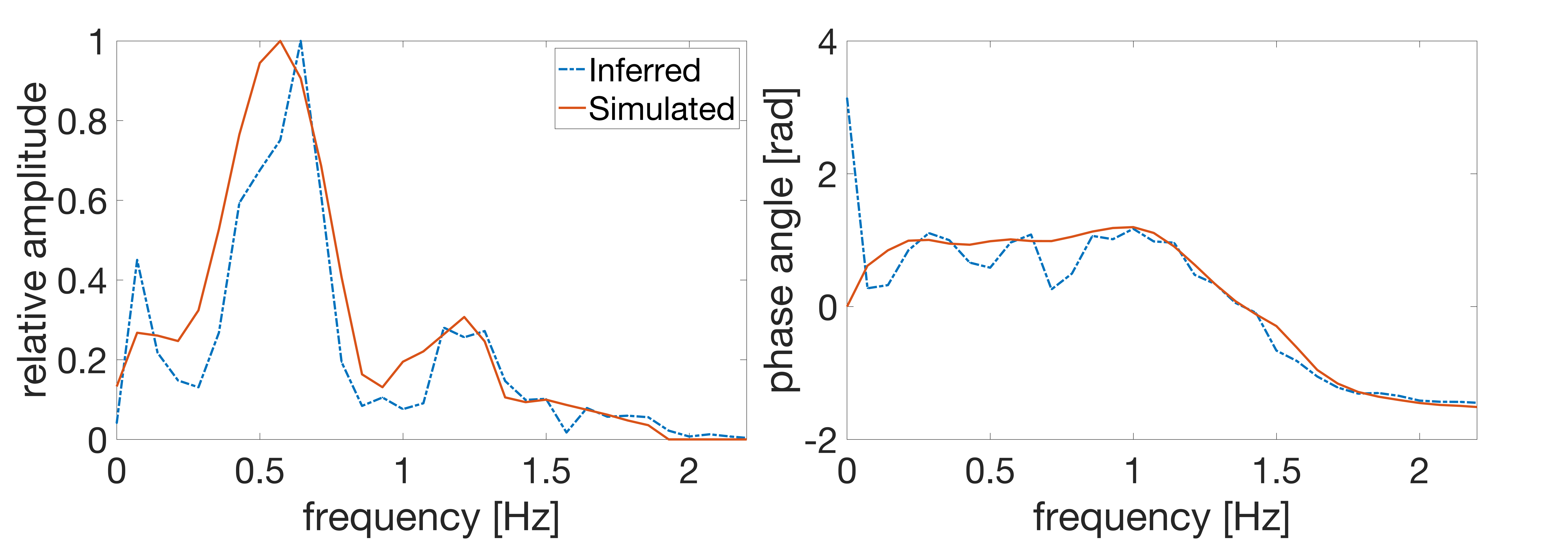}
    \caption{Spectrum of rotor speed $\omega_{13}$ response from $u_1$: simulated vs. inferred from ambient data.}
    \label{fig:spectrum_omega}
\end{figure}

\subsection{Single FO Source with Single FO Mode}

We first validated the algorithm using measurements of generator states. Despite violations of model assumptions, the Fourier-domain responses matched very well with the model-based responses in terms of amplitude and phase angle within the frequency band of interest. Figure~\ref{fig:spectrum_omega} compares the two responses for rotor speed $\omega$. Clearly, the match between the two is very close, except for some small deviations, primarily due to model approximations. By using the recovered responses, we were able to localize most FOs via the LS solutions in~\eqref{eq:fo_localization0}. {  Among all 64 scenarios, the only error of LS solutions comes from the less significant system mode 0.57Hz.} However, the source can still be recovered from the neighboring set of the identified source, showing the validity of the proposed model and algorithm. The experiment setup and test results are shown in Table~\ref{table:benchmark}.

Next, we simulated a scenario with measurements of the generators' terminal buses. As mentioned before, due to the power flow coupling, the approximation error between generator speed and terminal bus frequency manifests itself in estimated responses, and leads to errors in FO source recovery. The numerical tests suggest that despite the reduced accuracy in inferred responses, the proposed algorithm can still recover most FO sources correctly from LS solutions and their neighboring sets, as shown in Table~\ref{table:benchmark}. 

\begin{figure}[t]
     \centering
         \includegraphics[width=1.0\linewidth]{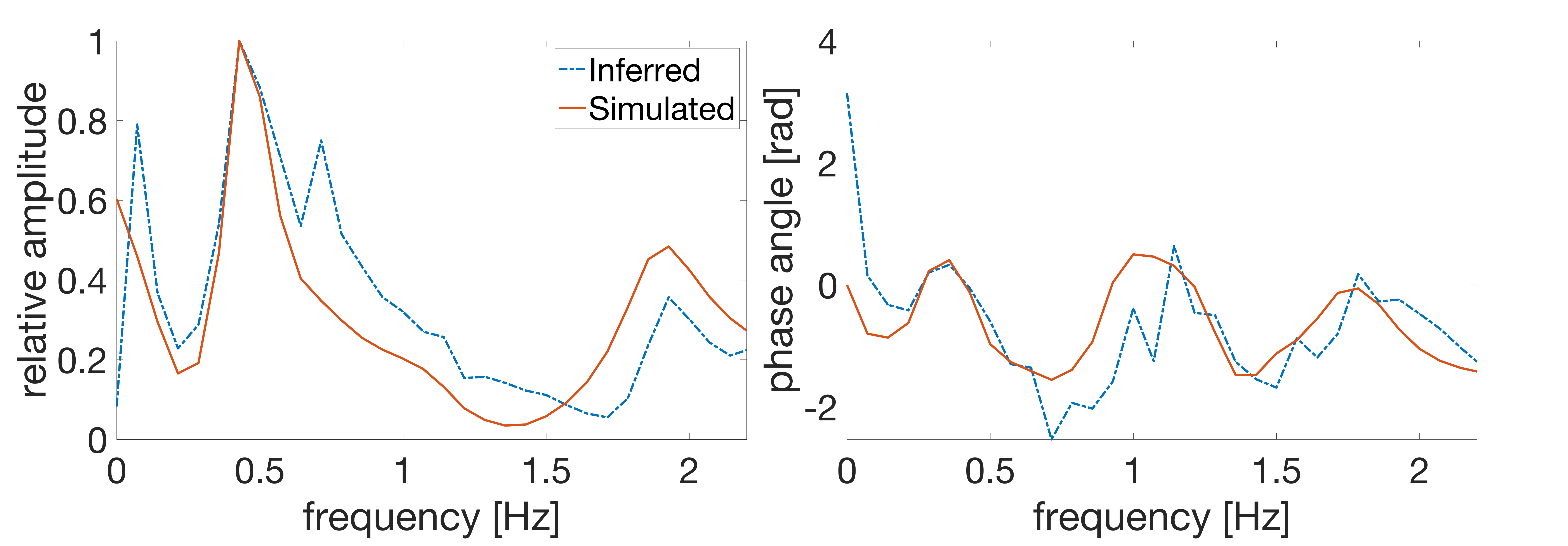}
    \caption{Spectrum of rotor speed $\omega_{13}$ (measured by $f_{37}$) response from $u_{11}$: simulated vs. inferred from ambient data. }
    \vspace*{-5pt}
    \label{fig:spectrum_theta}
\end{figure}

We also tested the proposed algorithm under partial sensor coverage with no measurements at generator buses. Each generator was surrogated by a measurement at a nearby bus. In the 68-bus system, the assumed PMUs were placed within at most 2 hops away as depicted in Fig.~\ref{fig:68bus_diagram}. We considered measurements of both types, bus frequency $f_i$ and line flow $p_{ij}$. As suggested in Section~\ref{sec:general}, relative phase shifts are used for FO localization when measurements for input locations are not near generators. The results advocate that FO sources recovered from LS solutions are less accurate due to inaccurate response estimations, as shown by the example in Fig.~\ref{fig:spectrum_theta}. The proposed algorithm still points out the correct FO source area, as the correct source can be recovered from the reported neighboring set in most cases, as reported in Table~\ref{table:benchmark}. 

\begin{table}[t]
\centering
\caption{Localization accuracy for single-source FOs}
\begin{tabular}{|l|r|r|r|}
\hline\hline
\textbf{Input} & \textbf{Sensor coverage} & \textbf{LS} & \textbf{Graph} \\
\hline\hline
Governor   & Rotor (Full)            & 98.40\%     & 100\%          \\
Governor   & Bus (Full)            & 92.20\%     & 96.90\%        \\
Governor   & Bus (Partial)         & 87.53\%     & 98.50\%        \\
Governor   & Bus\&Line (Partial)         & 82.80\%     & 92.20\%        \\
{  Exciter}    & {  Bus (Full)}         & {  30.60\%}     & {  97.20\%}   \\
{  Exciter}    & {  Bus (Partial)}         & {  36.10\%}     & {  94.50\%}   \\
\hline\hline    
\end{tabular}
\label{table:benchmark}
\end{table}
Finally, disturbances originating from the generator's excitation system were considered. {  We perturbed the voltage reference at 9 generators equipped with exciters to generate simulated impulse responses and FOs. For FOs, we simulated 4 identified modes, so there were 36 scenarios.} In this case, excitation system outputs are dominated by the dynamics of the excitation system. Thus, the responses of states and variables are delayed, and the linearized system may not reflect the actual responses well. Under the realistic partial sensor coverage setup, we noticed that the FO sources recovered from LS solutions were inaccurate. Nevertheless, the correct FO source area is still pointed out correctly in the neighboring set in most cases, as reported in Table~\ref{table:benchmark}.


\begin{table}[t]
\centering
\caption{Localization accuracy for 2 FO sources}
\begin{tabular}{|l|r|r|r|}
\hline\hline 
\textbf{Sensor coverage} & \textbf{LS} & \textbf{Graph} & \textbf{Major mode} \\
\hline\hline
Rotor (Full)            & 90.23\%     & 98.05\%   &   99.61\%     \\
Bus (Partial)            & 43.95\%     & 91.21\%   &  96.48\%    \\
\hline\hline  
\end{tabular}
\label{table:2loc_2mode}
\end{table}

\subsection{Two FO Sources with Multiple Modes}
We also considered the case of two FO sources at two distinct input frequencies. The input frequencies were selected at the identified system modes, with $0.143$Hz set as the major mode with twice the input amplitude as $0.43$Hz. This case is more challenging as the major mode may dominate system oscillations. As detailed in Table~\ref{table:2loc_2mode}, the proposed algorithm can still identify both FO frequencies and locate the sources correctly {  for almost all 256 possible scenarios}. With full observability on rotor states, Algorithm~\ref{alg:fo_framework}  missed the actual major FO source in a single scenario and successfully identified both sources for 251 scenarios. Under the more realistic condition with measurements only at proxy buses, Algorithm~\ref{alg:fo_framework} still achieves high accuracy in locating the major FO source. Both sources can be recovered from the reported neighboring set for the vast majority of scenarios, showing the effectiveness and robustness of the proposed FO localization algorithm.

\begin{figure}[t]
     \centering
         \includegraphics[width=1.0\linewidth]{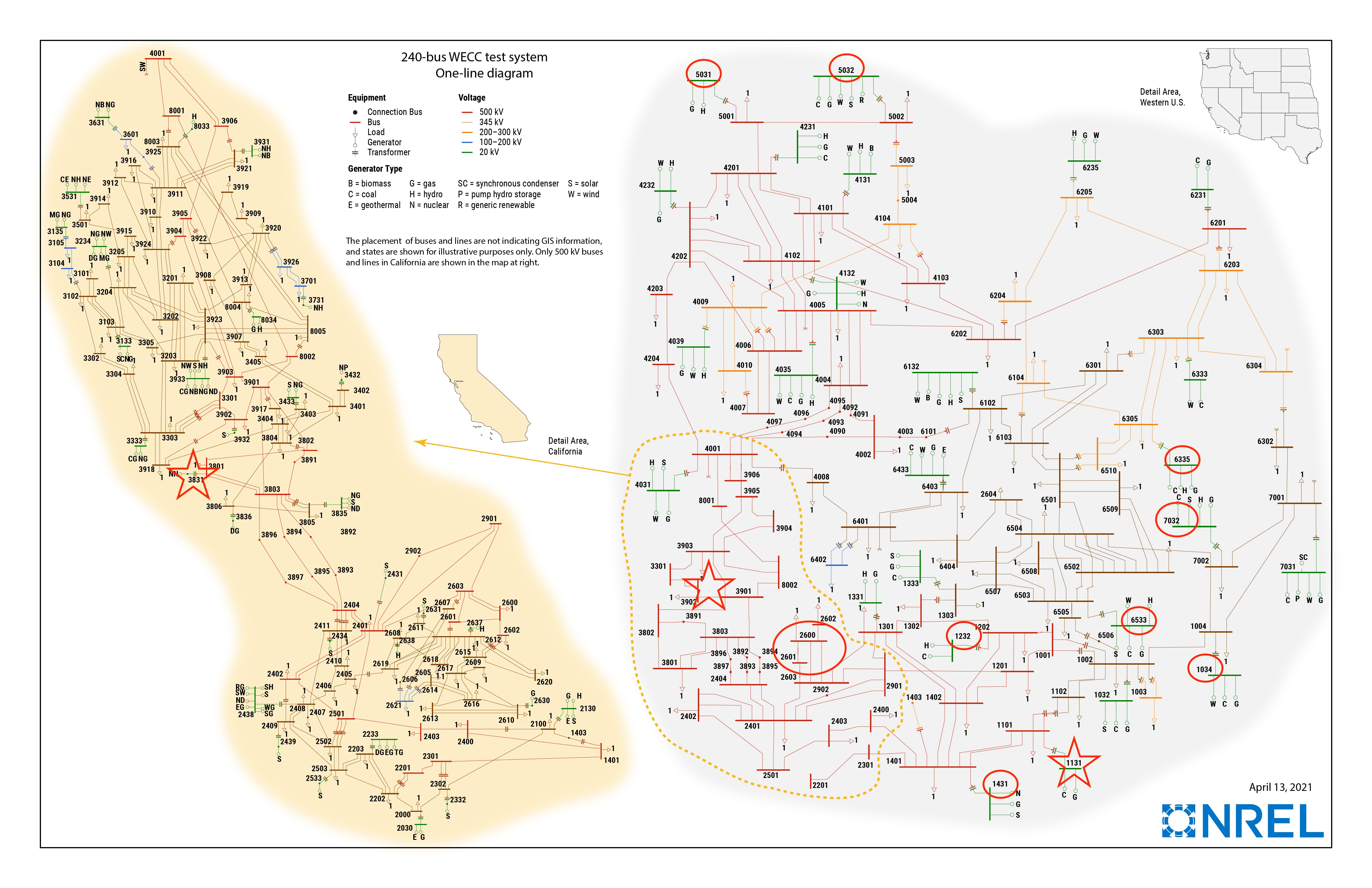}
    \caption{{  Diagram of the 240-bus system with 26 monitored generators~\cite{yuan2020developing}. Circled locations are monitored generators with both name and rating matched at the ambient and FO datasets. Locations with stars are FO sources in Case 3 and Case 4.}}
    \label{fig:240bus_diagram}
\end{figure}

\subsection{{  Validation with the IEEE-NASPI FO Contest Dataset}}

{  This 240-bus system was used as a benchmark by the FO localization contest~\cite{wang2021ieee}. The power system includes a total of 13 test cases. Here, we focus on Case 3 and Case 4, which are considered to be the most challenging cases with both resonance conditions and un-monitored sources. A 15-min ambient dataset was used to infer impulse responses, while FO signals lasting for 15 seconds from the contest dataset were subsequently used for the localization. Due to some mismatches in the naming and data types between the available ambient and FO datasets, we were only able to utilize the rotor frequency data at 9 out of the 26 measured generators to evaluate Algorithm~\ref{alg:fo_framework}, as marked in Fig.~\ref{fig:240bus_diagram}. For Case 3, Algorithm~\ref{alg:fo_framework} has successfully identified bus 1431 as the FO source regardless of the presence of fault-induced transients. Indeed, bus 1431 is the closest bus to bus 1131, the actual source, among all monitored locations. For the 3 scenarios under Case 4 with mostly pre-fault FO signals, Algorithm~\ref{alg:fo_framework} identified bus 2634, which is also the closest one to the actual source at bus 3831. If the post-fault FO signals were considered, the source was identified as  bus 1232, which is also very close to the actual source. This numerical test has shown promising results in using the proposed algorithm for a realistic large test system, even if the FO source is not directly monitored.
}

\section{Conclusions}
\label{sec:conclusion}
This work has developed a data-driven framework for FO localization with dynamics responses recovered from ambient synchrophasor data. FO responses have been analyzed for both generator states and output variables as buses and lines. Leveraging the recovered response, the FO localization task has been posed as an LS fitting problem. A computationally efficient algorithm has been proposed that uses ambient and oscillatory phasor measurements to identify FO sources. The algorithm is flexible regarding location and types of PMU data, and can provide a system operator with a neighboring set of possible FO locations. The proposed method has been numerically validated using the IEEE 68-bus system under realistic nonlinear dynamics with control effects, as well as a dataset on a 240-bus system generated for the IEEE-NASPI contest~\cite{wang2021ieee}. The method could pinpoint FO sources for the vast majority of scenarios, under full and partial observability, governor or exciter oscillations. {  Future research directions include model-based analysis for different types of FO sources, performance analysis for various measurement types and operation conditions, as well as numerical validation using real-world PMU data.}

 \section{Acknowledgements}
The authors would like to thank Dr.~Joe Chow for his valuable suggestions on the 68-bus system simulations in PST, as well as Dr.~Bin Wang and Ms.~Lingjuan Chen for kindly providing us the ambient dataset for the 240-bus system and for their helpful discussions on interpreting the dataset.

%
\bibliographystyle{IEEEtran}
\itemsep2pt
\bibliography{IEEEabrv,hzpub,ref,kekatos}

\begin{thebibliography}{10}
\providecommand{\url}[1]{#1}
\csname url@samestyle\endcsname
\providecommand{\newblock}{\relax}
\providecommand{\bibinfo}[2]{#2}
\providecommand{\BIBentrySTDinterwordspacing}{\spaceskip=0pt\relax}
\providecommand{\BIBentryALTinterwordstretchfactor}{4}
\providecommand{\BIBentryALTinterwordspacing}{\spaceskip=\fontdimen2\font plus
\BIBentryALTinterwordstretchfactor\fontdimen3\font minus
  \fontdimen4\font\relax}
\providecommand{\BIBforeignlanguage}[2]{{%
\expandafter\ifx\csname l@#1\endcsname\relax
\typeout{** WARNING: IEEEtran.bst: No hyphenation pattern has been}%
\typeout{** loaded for the language `#1'. Using the pattern for}%
\typeout{** the default language instead.}%
\else
\language=\csname l@#1\endcsname
\fi
#2}}
\providecommand{\BIBdecl}{\relax}
\BIBdecl

\bibitem{nerc2021oscillationref}
NERC, ``{Recommended Oscillation Analysis for Monitoring and Mitigation
  Reference Document},'' \emph{NERC: Atlanta, GA, USA}, Nov. 2021.

\bibitem{fo_report2023}
L.~Chen, D.~Trudnowski, L.~Dosiek, S.~Kamalasadan, Y.~Xu, and X.~Wang,
  ``{Forced Oscillations in Power Systems},'' \emph{IEEE TR110}, 2023.

\bibitem{MASLENNIKOV201755}
S.~Maslennikov, B.~Wang, and E.~Litvinov, ``{Dissipating energy flow method for
  locating the source of sustained oscillations},'' \emph{International Journal
  of Electrical Power \& Energy Systems}, vol.~88, pp. 55--62, 2017.

\bibitem{Osipov2022cpsd}
D.~Osipov, S.~Konstantinopoulos, and J.~H. Chow, ``{A Cross-Power Spectral
  Density Method for Locating Oscillation Sources using Synchrophasor
  Measurements},'' \emph{{IEEE} Trans. Power Syst.}, pp. 1--9, 2022.

\bibitem{chen2013def}
L.~Chen, Y.~Min, and W.~Hu, ``{An energy-based method for location of power
  system oscillation source},'' \emph{{IEEE} Trans. Power Syst.}, vol.~28,
  no.~2, pp. 828--836, 2013.

\bibitem{maslennikov2021}
S.~Maslennikov and E.~Litvinov, ``{ISO New England Experience in Locating the
  Source of Oscillations Online},'' \emph{{IEEE} Trans. Power Syst.}, vol.~36,
  no.~1, pp. 495--503, 2021.

\bibitem{lesieutre2022model}
B.~C. Lesieutre, Y.~Abdennadher, and S.~Roy, ``Model-enhanced localization of
  forced oscillation using pmu data,'' in \emph{Allerton Conf. on
  Communication, Control, and Computing}.\hskip 1em plus 0.5em minus
  0.4em\relax IEEE, 2022, pp. 1--8.

\bibitem{huang2020forced}
T.~Huang, N.~M. Freris, P.~R. Kumar, and L.~Xie, ``{A Synchrophasor Data-Driven
  Method for Forced Oscillation Localization Under Resonance Conditions},''
  \emph{{IEEE} Trans. Power Syst.}, vol.~35, no.~5, pp. 3927--3939, 2020.

\bibitem{delabays2022locating}
R.~Delabays, A.~Y. Lokhov, M.~Tyloo, and M.~Vuffray, ``Locating the source of
  forced oscillations in transmission power grids,'' \emph{{PRX Energy}},
  vol.~2, p. 023009, Jun 2023.

\bibitem{cai2022online}
Y.~Cai, X.~Wang, G.~Jo{\'o}s, and I.~Kamwa, ``An online data-driven method to
  locate forced oscillation sources from power plants based on sparse
  identification of nonlinear dynamics {(SINDy)},'' \emph{{IEEE} Trans. Power
  Syst.}, vol.~38, no.~3, pp. 2085--2099, 2022.

\bibitem{huynh2018data}
P.~Huynh, H.~Zhu, Q.~Chen, and A.~E. Elbanna, ``Data-driven estimation of
  frequency response from ambient synchrophasor measurements,'' \emph{{IEEE}
  Trans. Power Syst.}, vol.~33, no.~6, pp. 6590--6599, 2018.

\bibitem{liu2023dynamic}
S.~Liu, H.~Zhu, and V.~Kekatos, ``Dynamic response recovery using ambient
  synchrophasor data: A synthetic {Texas Interconnection} case study,'' in
  \emph{{Hawaii Intl. Conf. on System Sciences (HICSS)}}, Maui, HI, Jan. 2023.

\bibitem{arthur2000power}
R.~B. Arthur and V.~Vittal, ``{Power System Analysis},'' \emph{2nd ed., London:
  UK}, pp. 532--538, 2000.

\bibitem{chevalier2018mitigating}
S.~C. Chevalier and P.~D. Hines, ``Mitigating the risk of voltage collapse
  using statistical measures from {PMU} data,'' \emph{{IEEE} Trans. Power
  Syst.}, vol.~34, no.~1, pp. 120--128, 2018.

\bibitem{venkatasubramanian16resonance}
S.~A.~N. Sarmadi and V.~Venkatasubramanian, ``{Inter-Area Resonance in Power
  Systems From Forced Oscillations},'' \emph{{IEEE} Trans. Power Syst.},
  vol.~31, no.~1, pp. 378--386, 2016.

\bibitem{jalali2021inferring}
M.~Jalali, V.~Kekatos, S.~Bhela, and H.~Zhu, ``Inferring power system frequency
  oscillations using {G}aussian processes,'' in \emph{IEEE Conf. on Decision
  and Control}, Austin, TX, 2021, pp. 3670--3676.

\bibitem{Paganini19}
F.~{Paganini} and E.~{Mallada}, ``{Global Analysis of Synchronization
  Performance for Power Systems: Bridging the Theory-Practice Gap},''
  \emph{{IEEE} Trans. Autom. Control}, vol.~65, no.~7, pp. 3007--3022, 2020.

\bibitem{osti_1004165}
\BIBentryALTinterwordspacing
P.~Mackin, R.~Daschmans, B.~Williams, B.~Haney, R.~Hung, and J.~Ellis,
  ``{Dynamic Simulation Studies of the Frequency Response of the Three U.S.
  Interconnections with Increased Wind Generation},'' \emph{LBNL Technical
  Report}, Dec. 2010. [Online]. Available:
  \url{https://www.osti.gov/biblio/1004165}
\BIBentrySTDinterwordspacing

\bibitem{sauer2017power}
P.~W. Sauer, M.~A. Pai, and J.~H. Chow, \emph{Power system dynamics and
  stability: with synchrophasor measurement and power system toolbox}.\hskip
  1em plus 0.5em minus 0.4em\relax John Wiley \& Sons, 2017.

\bibitem{Trudnowski2008mode}
D.~J. Trudnowski, ``Estimating electromechanical mode shape from synchrophasor
  measurements,'' \emph{{IEEE} Trans. Power Syst.}, vol.~23, no.~3, pp.
  1188--1195, 2008.

\bibitem{hzgg_tps12}
H.~Zhu and G.~B. Giannakis, ``Sparse overcomplete representations for efficient
  identification of power line outages,'' \emph{{IEEE} Trans. Power Syst.},
  vol.~27, no.~4, pp. 2215--2224, 2012.

\bibitem{hale2006efficient}
D.~Hale, ``An efficient method for computing local cross-correlations of
  multi-dimensional signals,'' \emph{CWP Report}, vol. 656, 2006.

\bibitem{wang2021ieee}
B.~Wang and S.~Maslennikov, ``{IEEE-NASPI} oscillation source location
  contest-case development and results,'' National Renewable Energy Lab.(NREL),
  Golden, CO (United States), Tech. Rep., 2021.

\bibitem{chow1992pst}
J.~Chow and K.~Cheung, ``A toolbox for power system dynamics and control
  engineering education and research,'' \emph{{IEEE} Trans. Power Syst.},
  vol.~7, no.~4, pp. 1559--1564, 1992.

\bibitem{yuan2020developing}
H.~Yuan, R.~S. Biswas, J.~Tan, and Y.~Zhang, ``Developing a reduced 240-bus
  wecc dynamic model for frequency response study of high renewable
  integration,'' in \emph{2020 IEEE/PES transmission and distribution
  conference and exposition (T\&D)}.\hskip 1em plus 0.5em minus 0.4em\relax
  IEEE, 2020, pp. 1--5.

\end{thebibliography}
\end{document}